\documentclass[12pt]{article}
\usepackage{geometry}
\usepackage{a4}
\usepackage{graphicx,subfigure}
\usepackage{epsf}
\usepackage{amsmath}
\usepackage{amssymb}
\usepackage{cite}
\usepackage{tensor}
\begin{document}
\def\rh{r_H}
\def\rt{r_t}
\def\onefour{\frac{1}{4}}
\def\half{\frac{1}{2}}
\def\threefour{\frac{3}{4}}
\def\fivefour{\frac{5}{4}}
\def\sevenfour{\frac{7}{4}}
\def\mfour{\frac{m}{4}}
\def\nfour{\frac{n}{4}}
\def\s{{\mathbb{S}}}
\def\T{{\mathbb{T}}}
\def\Z{{\mathbb{Z}}}
\def\W{{\mathbb{W}}}
\def\Bbb{\mathbb}
\def\BZ{\Bbb Z} \def\BR{\Bbb R}
\def\BW{\Bbb W}
\def\BM{\Bbb M}
\def\BC{\Bbb C} \def\BP{\Bbb P}
\def\CP{\BC\BP}

\begin{titlepage}
\title{A Note on Subregion Holographic Complexity}
\author{}
\date{
 Pratim Roy, Tapobrata Sarkar
\thanks{\noindent E-mail:~ proy, tapo@iitk.ac.in}
\vskip0.4cm
{\sl Department of Physics, \\
Indian Institute of Technology,\\
Kanpur 208016, \\
India}}
\maketitle\abstract{
\noindent
The volume inside a Ryu-Takayanagi surface has been conjectured to be related to the complexity of subregions of the
 boundary field theory. Here, we study the behaviour of this volume analytically, when the entangling surface has a strip geometry.
We perform systematic expansions in the low and high temperature regimes for AdS-Schwarzschild and RN-AdS black holes. 
In the latter regime, we point out spurious divergences that
might occur due to the limitations of a near horizon expansion. A similar analysis is performed for extremal black holes, and 
at large charge, we find that there might be some new features of the volume as compared to the area. 
Finally, we numerically study a four dimensional 
RN-AdS black hole in global AdS, the entangling surface being a sphere. We find that the holographic complexity 
captures essentially the same information as the entanglement entropy, as far as phase transitions are concerned.}
\end{titlepage}

\section{Introduction}

The Ryu-Takayanagi (RT) conjecture \cite{RT} and its covariant generalisation are both proposals to calculate the holographic entanglement 
entropy of field theories in the context of the AdS/CFT correspondence, 
and have been used extensively in the recent past in several contexts. Broadly, the proposal computes holographic entanglement 
entropy of a region $A$ in a $d$ dimensional boundary by the formula
\begin{equation}
S_{EE} = \frac{Area(\gamma_A)}{4G_{d+1}}
\end{equation}
where $\gamma_A$ is the extremal surface in the bulk, that extends from the boundary of the region $A$, and $G_{d+1}$ is Newton's constant 
in the $(d+1)$ dimensional bulk. Entanglement being a central theme in theories of quantum information theory, the RT conjecture has spurred a lot 
of activity over the past decade, and has led to new insights into the nature of strongly coupled matter. 
 
Our focus in this paper is going to be another information theoretic quantity, called the holographic complexity. The 
complexity of a quantum state is roughly defined as the difficulty, or the number of steps required to prepare this state from a given reference
state. Recently, there have been two proposals to calculate the complexity of a quantum state from the holographic perspective. The 
first, popularly called the complexity = volume (CV) conjecture \cite{Susskind} sates that the complexity of a given state on the boundary CFT is dual to
the volume of the maximal hypersurface in the bulk of codimension one, that is matched to the boundary at a given time. The second, 
called the complexity = action (CA) conjecture relates the complexity to the gravitational action computed on the Wheeler-De Witt patch
\cite{SusskindA}, \cite{Couch}.

Associated to the CV conjecture is the notion of sub-region complexity, one that is related to the volume of the bulk enclosed by a
Ryu-Takayanagi surface (we will call this the RT volume in sequel). These are in particular related to specific subregions on a time slice at 
the boundary (i.e related to mixed states), and the proposal of \cite{Alishahiha} states that the holographic dual of the complexity in
such states is 
\begin{equation}
\mathcal{C} = \frac{V(\gamma_A)}{G_{d+1} L_{AdS}}
\label{volumemain}
\end{equation}
This definition admits an ambiguity up to a constant factor. Previous work on this quantity was reported in \cite{Omar} where
the authors studied some aspects of this volume in various contexts, including the pure AdS and AdS-Schwarzschild (SAdS) black holes. 
In a more recent work, \cite{Myers} have commented upon the generic singularity structure of the volume, in various scenarios. 

Our main aim in this paper is to understand the behaviour of sub-region complexity analytically, for $d$ dimensional AdS black holes with
planar horizon. We will
consider time-independent geometries, and our focus would be to systematically analyse the high temperature and low temperature 
expansions of the RT volume for various background geometries. In particular, we consider the SAdS, the 
Reissner-Nordstrom-AdS (RN-AdS) and extremal black holes in various dimensions. Intuitively it is clear that up to possible additive and
multiplicative numerical constants and possible functional dependence on some system parameters, at low temperatures
the finite part of the volume should scale as $V \sim T^{\alpha} + T^{2\alpha} + \cdots$, while at high temperatures it should behave like
$V \sim T^{\alpha}(1 + 1/T^{\beta})$, where $\alpha$ and $\beta$ are generic dimension dependent constants. The singular part of the
volume is expected to have a single divergence in terms of an ultra violet cutoff. In this work, we establish
such behaviour and we will see that in some cases, it leads to interesting deviations from expected results. 
Such analysis has not been performed elsewhere, and should be interesting as it is expected to provide further information about 
the boundary field theory. 

This paper is organised as follows. 
In the first part of this paper, in section 2, after setting up the necessary notations, we will develop a systematic expansion of the RT volume for
SAdS black holes. We will show that the behaviour of the volume at high temperatures is of the same form as that of the entanglement
entropy, modulo numerical factors, and the same is true for low temperatures as well. We then comment upon the low temperature
expansion of the volume, and find similar features. We point out in this part some spurious divergences that might occur in the volume,
if we resort to a near horizon approximation. 

In section 3, we move to the RN-AdS black holes, and show that such spurious divergences
will be present in that case as well, in the high temperature limit, while the behaviour of the RT volume in the low temperature limit mimics that
of the EE. Finally, we study the case of extremal black holes, and while in the low charge limits the behaviour of the volume is similar
to that of the entanglement entropy, we find that the large charge case has some new features compared to the EE. All these computations
are done with the entangling surface having a strip geometry. In the last part of this paper, in section 4, we comment on the volume 
with a spherical entangling surface. Working in a canonical ensemble, we show that the RT volume captures essentially the same
information as the EE. The paper ends with conclusions and a summary of results, in section 5. 

\section{Planar AdS Schwarzschild Black Holes}

In this section, we will consider holographic complexity, i.e the RT volume with a strip geometry, 
for AdS Schwarzschild (SAdS) black holes at high and low temperatures. 
We will first give a brief description of the set-up used to calculate holographic complexity for the conformal field theory dual to 
$AdS_{d+1}$. We consider a subsystem in the boundary theory in the form of a strip, which length $\ell$. The extent of the subsystem in the 
other spatial boundary coordinates is $L$. 

Let us consider the metric of a planar black hole in $AdS_{d+1}$,
\begin{equation}
ds^2 = -\frac{r^2}{L_{AdS}^2}f(r)dt^2 + \frac{L_{AdS}^2}{r^2 f(r)}dr^2 + \frac{r^2}{L_{AdS}^2}d \vec{x}^2
\label{sads}
\end{equation}
where for SAdS black holes, we have $f(r) = 1 - r_H^d/r^d$, with $r_H$ being the horizon radius of the black hole. 
Here, $L_{AdS}$ is the AdS length scale which is set equal to unity in the calculations that follow. 
The vector $\vec{x}$ corresponds to the boundary spatial coordinates, $\big\{ x,y_{1},y_{2},\ldots ,y_{d-2} \big\}$. 
As mentioned earlier, the boundary subsystem is chosen as $x \in \big[ -\frac{\ell}{2},\frac{\ell}{2} \big]$ 
and $\vec{y} \in \big[ -\frac{L}{2},\frac{L}{2} \big]$. Since the strip has translation invariance along $\vec{y}$, 
we can describe the profile of the extremal surface by $x \equiv x(r)$. For this set-up, the volume enclosed by the 
Ryu-Takayanagi surface extending from the boundary into the bulk is given by 
\begin{equation}
V = 2 L^{d-2} \int_{r_t}^{\infty } \frac{r^{d-2}
   x(r)}{\sqrt{f(r)}} \, dr
\end{equation}
Making a variable transformation, $r=\frac{r_t}{u}$, in terms of the new variable $u$ and after solving for $x(r)$, 
the expressions for the volume $V$ and the system size $l$ can be straightforwardly shown to be \cite{Omar}
\begin{eqnarray}
l &=& \frac{2}{r_t}\int_0^1 \frac{u^{d-1}}{\sqrt{1-u^{2d-2}}}\frac{1}{\sqrt{f(u)}} du\nonumber\\
V &=& 2L^{d-2}r_t^{d-2} \int_{\delta}^{1}\frac{1}{u^d\sqrt{f(u)}}\int_u^1 \frac{u^{d-1}}{\sqrt{1-u^{2d-2}}}\frac{1}{\sqrt{f(u)}} du
\label{mainvl}
\end{eqnarray}
We also record the expression for the RT surface, and this is given by 
\begin{equation}
A = 2L^{d-2}r_t^{d-2} \int_{\delta}^{1} \frac{1}{u^{d-1}{\sqrt{1-u^{2d-2}}}}\frac{1}{\sqrt{f(u)}} du
\label{maina}
\end{equation}

Here, $\delta$ is an ultraviolet cutoff, introduced to render the volume integral finite. 
For example, in the pure AdS case, one finds, by setting $f(u)=1$ that 
\begin{eqnarray}
l &=& \frac{2 \sqrt{\pi } \Gamma \left(\frac{d}{2 d-2}\right)}{r_t \Gamma
   \left(\frac{1}{2 (d-1)}\right)}\nonumber\\
V &=& 2L^{d-2}r_t^{d-2}\left(\frac{ \sqrt{\pi } \delta ^{1-d} \Gamma \left(\frac{d}{2
   d-2}\right)}{(d-1) \Gamma \left(\frac{1}{2 d-2}\right)} -\frac{2^{\frac{1}{d-1}} \Gamma \left(\frac{2 d-1}{2
   d-2}\right)^2}{\Gamma \left(\frac{1}{d-1}\right)} + \frac{ \delta}{d} +{\mathcal O}(\delta^{2d-1})\right)
\label{lvpureads}
\end{eqnarray}
where we have performed an expansion up to first order in $\delta$. 
This is the familiar statement that the divergence of the RT volume $\sim \delta^{1-d}$. Since we are dealing with a strip geometry, there
are no other divergences. It is useful to compare this with the entanglement entropy
or equivalently the area of the RT surface, given by 
\begin{equation}
A = L^{d-2}r_t^{d-2}\left(\frac{2\delta ^{2-d}}{d-2}+\frac{\sqrt{\pi } 
   \Gamma \left(-\frac{d-2}{2 (d-1)}\right)}{(d-1) \Gamma
   \left(\frac{1}{2 d-2}\right)} + {\mathcal O}(\delta^{3d-2})\right)
\label{EEpureads}
\end{equation}
Comparing eqs.(\ref{lvpureads}) and (\ref{EEpureads}), it is seen that modulo numerical factors, the divergences differs by a factor of $\delta^{-1}$, 
and that up to numerical factors, the finite part of both $V$ and $A$ $\sim (L/l)^{d-2}$. This was for pure AdS, and similar computations need to
be performed to determine $V$ and $l$ for the SAdS black hole. 
In principle, the procedure to compute these are simple, and for the cases we consider here, apart from the binomial expansion formula, 
we will only have to use the identity
\begin{equation}
\frac{1}{\sqrt{1-x}}= \sum_{n=0}^{\infty}\frac{1}{\sqrt{\pi}}\frac{\Gamma(n+\frac{1}{2})}{\Gamma(n+1)}x^n~,
\label{masteridentities}
\end{equation} 
to obtain a systematic expansion for $f(u)^{-1/2}$, with $x \equiv r_H/r_t$. Such an analysis for the entanglement entropy 
and some other non-local quantities was performed in \cite{Sandipan}. It is well known \cite{Hubeny} that $r_H < r_t$ for any static,
spherically symmetric space time, i.e the turning point of the extremal surface cannot penetrate the black hole horizon. Hence, we will
have two cases, $x \ll 1$ and $x \to 1$, corresponding to low and high temperatures, respectively, or low and high charge respectively,
for extremal black holes. In the former case, one can solve for $l$ of eq.(\ref{mainvl})
perturbatively, in terms of $r_t$. This can then be used to obtain $A$ of eq.(\ref{maina}) as a function of $l$. In the other limit, i.e when
$x \to 1$, the series for $l$ and $A$ are both formal divergences. In that case, using gamma function identities, 
one can always rewrite the series in $A$ as a part proportional to $l$ and a finite contribution \cite{Sandipan}. 
Hence, one can again obtain $A$ as a function of $l$. 

The procedure of \cite{Sandipan}, translated in our case is then to obtain systematic expansions for the RT volume $V$. 
Once such an expansion is done, we can identify these expansions with those of $l$ and hence obtain $V$ as a function of $l$. 
The only complication in this case arises as we have a double 
summation in the formula for $V$ (eq.(\ref{mainvl})) compared to a single sum in eq.(\ref{maina}), the latter being substantially easier to handle. 
Let us now illustrate this explicitly with the example of the SAdS metric of eq.(\ref{sads}). 

We will first record the expression for $l$, and this can be shown to be given by 
\begin{equation}
l = \frac{1}{r_t}\sum_{n=0}^{\infty}\left(\frac{r_H}{r_t}\right)^{d n}\frac{\Gamma \left(n+\frac{1}{2}\right) \Gamma \left(\frac{d (n+1)}{2
   (d-1)}\right)}{(d-1)  \Gamma (n+1) \Gamma \left(\frac{d (n+2)-1}{2
   (d-1)}\right)}
\label{lsads}
\end{equation}
For large $n$, $l \sim \sum_n x^{dn}/n$, and the divergent part in the limit of high temperature, i.e $r_H \to r_t$ ($x\to 1$ )
$\sim \log(1 - (r_H/r_t)^d)$. From our previous discussion, we will need to compute the RT volume, and relate its divergence at
high temperatures to that of $l$. For the low temperature case, the issue is much simpler, as we will see in sequel. 

For the SAdS black hole, the volume expanded in powers of $\delta$, turns out to be
\begin{equation}
V = L^{d-2}r_t^{d-2}\sum_{m=0}^{\infty}\sum_{n=0}^{\infty} \left(\frac{r_H}{r_t}\right){}^{d (m+n)}\left(T_1 + T_2 + T_3 \cdots\right)
\label{vol1}
\end{equation}
where we have
\begin{eqnarray}
T_1 &=& \frac{2 \Gamma \left(m+\frac{1}{2}\right) \Gamma
   \left(n+\frac{1}{2}\right) \delta ^{d m+d n+1}}{\pi  d \Gamma (m+2)
   \Gamma (n+1) (d m+d n+1)}~,\nonumber\\
T_2 &=& -\frac{\Gamma \left(m+\frac{1}{2}\right) \Gamma
   \left(n+\frac{1}{2}\right) \Gamma \left(\frac{dm+d}{2 d-2}\right) \delta ^{-d+d n+1}}{\sqrt{\pi } (d-1) (d n-d+1) \Gamma
   (m+1) \Gamma (n+1) \Gamma \left(\frac{dm + 2 d-1}{2 d-2}\right)}~,\nonumber\\
T_3 &=& \frac{\Gamma \left(m+\frac{1}{2}\right) \Gamma
   \left(n+\frac{1}{2}\right) \Gamma \left(\frac{d m + d n + 1}{2 d-2}\right)}{(d-1) \sqrt{\pi } (d n-d+1) \Gamma
   (m+1) \Gamma (n+1) \Gamma \left(\frac{d m + dn + d}{2 d-2}\right)}
\label{termssads}
\end{eqnarray}
and the $``\cdots"$ in eq.(\ref{vol1}) denote possible ${\mathcal O}(\delta^2)$ and higher terms, which we will ignore. Let us analyse the terms in eq.(\ref{termssads})
in some detail. These are all double sums, and one has to be careful about their convergence properties, in the limit $r_H \to r_t$. We will use the standard comparison
test for double sums (see, e.g chapter 7 of \cite{Ghorpade}) : say we have a sequence of real numbers $x(m,n)$ and $y(m,n)$, where $(m,n)$ 
are natural numbers (positive integers, in our case). Then, say $|x(m,n)| \leq |y(m,n)|$, for all $m$ and $n$. Then the comparison test implies that if 
$\sum_{m,n} y(m,n)$ is convergent, $\sum_{m,n} x(m,n)$ converges absolutely, and the latter is numerically less than the former, in absolute value. 

We will now discuss the limit $r_H \to r_t$ in some details. 
In order to check convergence of the series in eq.(\ref{termssads}), we will identify the behaviour of these in the limit
of large $m$ and $n$, and extract any divergence that might result. To apply the 
comparison test in our case, where we will choose $y(m,n) = m^{-3/2}n^{-3/2}$, with the known result that 
$\sum_{m,n=1}^{\infty} m^{-3/2}n^{-3/2} = \zeta \left(\frac{3}{2}\right)^2$, where $\zeta$ is the Riemann zeta function. Consider, for example,
the series involving $T_1$. In the limit of large $m$ and $n$, we have
\begin{equation}
T_1 \sim m^{-3/2}n^{-1/2}\left(3m +3 n\right)^{-1} \left(\frac{\delta r_H}{r_t}\right) < m^{-3/2}n^{-3/2}~,~~
\end{equation}
for all positive integer doublets $(m,n)$. The series involving $T_1$ thus converges absolutely, and since it has a multiplicative factor
of $\delta$, this term can be ignored. We are thus left with the terms $T_2$ and $T_3$ of eq.(\ref{termssads}). The asymptotic behaviour of
$T_2 \sim n^{-3/2}m^{-1/2}$. It is seen that the comparison test mentioned above fails in this case, and the sum over $m$ is
divergent. However, this is easy to understand. We first perform the $n$-sum to obtain 
\begin{equation}
\sum_{n=0}^{\infty}T'_2= \left(\frac{r_H}{r_t}\right)^{d m}\frac{\delta ^{1-d} \Gamma \left(\frac{1}{2} (2 m+1)\right) \Gamma
   \left(\frac{m d+d}{2 (d-1)}\right) \,
   _2F_1\left(\frac{1}{2},\frac{1}{d}-1;\frac{1}{d};\delta ^d
   \left(\frac{r_H}{r_t}\right){}^d\right)}{(d-1)^2 \Gamma (m+1) \Gamma
   \left(\frac{m d+2 d-1}{2 (d-1)}\right)}
   \end{equation}
where we have denoted $T'_2 = (r_H/r_t)^{dn + dm}T_2$. This can be expanded in powers of $\delta$, and the leading behaviour is 
\begin{eqnarray}
\sum_{n=0}^{\infty}\sum_{m=0}^{\infty}T'_2=
\sum_{m=0}^{\infty}\left(\frac{r_H}{r_t}\right)^{d m}\left(\frac{ \delta ^{1-d} \Gamma \left(m+\frac{1}{2}\right) \Gamma
   \left(\frac{d (m+1)}{2 (d-1)}\right)}{(d-1)^2 \Gamma (m+1) \Gamma
   \left(\frac{d (m+2)-1}{2 (d-1)}\right)}\right) \nonumber\\
   -\sum_{m=0}^{\infty}\left(\frac{r_H}{r_t}\right)^{d m+1}\left(\frac{(4 d+4) \delta  \Gamma \left(m+\frac{1}{2}\right) \Gamma
   \left(\frac{d (m+1)}{2 (d-1)}\right)
   }{8 (d-1) (d+1) \Gamma (m+1) \Gamma
   \left(\frac{d (m+2)-1}{2 (d-1)}\right)}\right)
   \end{eqnarray}
Hence, comparing with eq.(\ref{lsads}) we have in the limit $r_H \to r_t$,
\begin{equation}
\sum_{m=0}^{\infty}\sum_{n=0}^{\infty}T'_2\sim\left(\frac{a_1}{\delta ^{d-1}}+a_2\delta\right)lr_t
 \end{equation}
where $a_1$ and $a_2$ are ${\mathcal O}(1)$ constants that can be determined from the above. 
Finally, we are left with $T_3$, which is slightly more complicated to analyse. In this case, it is better to specialise to a given value of $d$, 
which we choose to be $d=3$. Then, we note that for a given finite value of $n$, the $m$-sum diverges.
Hence, we first sum over $n$, which yields a complicated formula in terms of generalised Hypergeometric functions, multiplied
by gamma functions. In order to extract the 
leading divergent behaviour, we evaluate the limiting form of these generalised Hypergeometric functions for large $m$. Then, upon 
using the limiting forms of the gamma functions for large arguments, we find 
\begin{equation}
\sum_{n=0}^{\infty}\left(\frac{r_H}{r_t}\right)^{3m+3n}T_3 \sim \left(\frac{r_H}{r_t}\right)^{3m} m^{-1},~~~m \gg 1 
\end{equation}
In the limit $r_H \to r_t$ therefore, these will yield the same divergence as $l$ of eq.(\ref{lsads}), 
and putting everything together, we obtain 
\begin{equation}
\sum_{m=0}^{\infty}\sum_{n=0}^{\infty}\left(\frac{r_H}{r_t}\right)^{3m+3n}T_3 \sim  lr_t\left(a_3 + \frac{a_4}{lr_t}\right),~~~(d=3)
\end{equation}
where $a_3$ and $a_4$ are ${\mathcal O}(1)$ constants, with $a_4$ being the finite part of the series over $T_3$ (we will momentarily
come back to this).
 Although the above equation is for $d=3$, we expect similar results for any other dimension, only
that the algebra becomes more complicated for higher dimensions. Hence, we record our final expression for the RT volume in the
limit $r_H \to r_t$ (with $a_i = {\mathcal O}(1)$) :
\begin{equation}
V \sim \left(\frac{a_1}{\delta ^{d-1}}+a_2\delta + a_3 + \frac{a_4}{lr_t}\right)lL^{d-2}r_t^{d-1}
\label{finalvsads}
\end{equation}
The divergent part arises from the pure AdS, and the renormalised volume is simply obtained by subtracting the contribution of pure AdS
from eq.(\ref{finalvsads}). 
It is useful to compare this calculation with the well known one for the entanglement entropy, or the area of the Ryu-Takayanagi surface,
given by 
\begin{eqnarray}
A &=& L^{d-2}r_t^{d-2}\sum_{n=0}^{\infty}\left(\frac{r_H}{r_t}\right)^{nd}{\mathcal F}(n)~,\nonumber\\
{\mathcal F}(n) &=& 
-\frac{\Gamma \left(n+\frac{1}{2}\right) \delta ^{d (n-1)+2} \Gamma
   \left(\frac{d (n-1)+2}{2 (d-1)}\right)}{\sqrt{\pi } (d-1) \Gamma
   (n+1) \Gamma \left(\frac{d (n+1)}{2 (d-1)}\right)} + 
   \frac{\Gamma \left(n+\frac{1}{2}\right) \Gamma \left(\frac{d (n-1)+2}{2
   (d-1)}\right)}{(d-1) \Gamma (n+1) \Gamma \left(\frac{d n+1}{2
   (d-1)}\right)}
\label{EESAdS}
\end{eqnarray}
In the limit $r_H \to r_t$, the second term in eq.(\ref{EESAdS}) is divergent, and by a slight rewriting of the gamma functions, this term
be shown to be equivalent to a term proportional to $l$ of eq.(\ref{lsads}) (which is formally divergent in this limit) along with a convergent piece. 
In our case for the RT volume, an essentially similar thing happens, but due to the complicated nature of the double sum over $T_3$ of 
eq.(\ref{termssads}), the finite part could not be written explicitly. In any case, the RT area, following the arguments above reduce to
\begin{equation}
A \sim \frac{b_1L^{d-2}r_t^{d-2}}{\delta ^{d-2}} + \left(b_2 + \frac{b_3}{lr_t}\right)lL^{d-2}r_t^{d-1}
\end{equation}
with the divergent term coming from $n=0$ in eq.(\ref{EESAdS}). The finite part is, with ${\mathcal V} = L^{d-2}l$ being the volume of the
rectangular strip being considered, and $T \sim r_H \to r_t$,
\begin{equation}
A_{finite} = \left( b_2 + \frac{b_3}{lr_t}\right)lL^{d-2}r_t^{d-1} \sim {\mathcal V}T^{d-1}\left(1 + {\mathcal O}\left(\frac{1}{Tl}\right)\right)
\end{equation}
Hence, from eq.(\ref{finalvsads}), we obtain a similar result, 
\begin{equation}
V_{finite} \sim {\mathcal V}T^{d-1}\left(1 + {\mathcal O}\left(\frac{1}{Tl}\right)\right)
\end{equation}

Let us now discuss the limit $r_H \ll r_t$, i.e the low temperature limit. This is simple to do, and we will be brief here. For simplicity, we will
work out the case $d=3$, although it should be obvious to the reader that the qualitative features of our analysis will not change for any $d$. 
We define $\epsilon = r_H/r_t$, and in terms of this variable, it can be shown from eq.(\ref{lsads}) that 
\begin{equation}
l_{d=3} = \frac{1}{r_t}\left(\frac{\pi  \epsilon^3}{8}+\frac{2 \sqrt{\pi } \Gamma
   \left(\frac{3}{4}\right)}{\Gamma \left(\frac{1}{4}\right)} + {\mathcal O}(\epsilon^6)\right)
\label{llow}
\end{equation}
This can be solved for $r_t$, and used in the equation for the complexity, which we now compute for low temperatures from eq.(\ref{termssads}). 
In that equation, we take the terms $T_2$ and $T_3$, compute the series with $r_H/r_t=\epsilon$, and expand the resulting Hypergeometric 
functions around $\epsilon = 0$. After collecting all the terms, we arrive at 
\begin{equation}
V_{d=3} = Lr_t\left(\frac{\pi  \epsilon ^3}{16 \delta ^2}+\frac{\Gamma
   \left(-\frac{1}{4}\right)^2}{16 \sqrt{2 \pi } \delta
   ^2}-\frac{\Gamma \left(\frac{1}{4}\right)^2}{4
   \sqrt{2 \pi }}+\frac{\epsilon ^3}{4}\right)
\label{vlow}
\end{equation}
We note that the first term in this sum $\sim \epsilon^3/\delta^2$ and might become indeterminate in the limit of $\epsilon \to 0$ and $\delta \to 0$,
and that this limit has to be taken carefully. 
For comparison, we record here the expression for the RT area in this limit for $d=3$\footnote{If we choose $\delta = r_t/r_b$, where the 
UV divergence is as $r_b \to 0$, then the divergent term in the area is universal. However, this may not generically be so.}
\begin{equation}
A_{d=3} =  Lr_t\left(\frac{\sqrt{\pi}\Gamma\left(-\frac{1}{4}\right)}{2\Gamma\left(\frac{1}{4}\right)} 
- \frac{\Gamma\left(-\frac{1}{4}\right)}{2\delta \Gamma\left(\frac{3}{4}\right)}+ \frac{\pi\epsilon^3}{4}\right)
\label{EElowT}
\end{equation}
At leading order, from eq.(\ref{llow}), $lr_t \sim {\mathcal O}(1)$. Hence, we obtain (remembering $T \sim r_H$),
\begin{equation}
A_{finite,d=3} = \left(\frac{L}{l}\right)\left(c_1 + {\mathcal O}(Tl)^3\right)~,~~(d=3)
\end{equation}
which translates to, for a general $d$, 
\begin{equation}
A_{finite} = \left(\frac{L}{l}\right)^{d-2}\left(c_1 + {\mathcal O}(Tl)^d\right)
\end{equation}
For the subregion complexity, we similarly obtain, generalising eqs.(\ref{vlow}) and (\ref{llow}),
\begin{equation}
V_{finite} = \left(\frac{L}{l}\right)^{d-2}\left(c_2 + {\mathcal O}(Tl)^d\right)
\end{equation}
In the above, $c_i$ are ${\mathcal O}(1)$ numerical factors. 

Before we proceed, we should point out that the SAdS case can in principle be done analytically for all values of the horizon radius
in the sense that the terms in eq.(\ref{termssads}) have been obtained exactly, starting from eq.(\ref{sads}). However, in 
the presence of a chemical potential, the situation will be more complicated, as we will illustrate below. In those cases, one has to resort
to a near horizon expansion in the high temperature limit (or, for extremal black holes in the large charge limit). This essentially requires
truncating the series for $f(u)$ up to a certain order in $1 - r_H/r_t$. In that case, spurious divergences in addition to the ones discussed
above might appear, which are artefacts of the near horizon expansion. To illustrate this, 
let us set up a computation of the RT volume in the SAdS case at high temperatures, with the near horizon approximation $r_H \to r_t$,
and see what are the qualitative differences that arise compared to the exact calculation given above, in that limit. 
Since this is the only purpose of the calculation below, we might as well simplify things by using the example of $d=3$.

In $d=3$, in terms of $\alpha = \frac{r_t}{r_H} - u$, we write
\begin{equation}
f(r) = 1 - \frac{r_H^3}{r^3} = 3\frac{r_H}{r_t}\alpha - 3\frac{r_H^2}{r_t^2}\alpha^2  - \frac{r_H^3}{r_t^3}\alpha^3
\label{sadsnhmetric}
\end{equation}
In the near horizon approximation, we will retain the first term in the above series.\footnote{In $d$-dimensions, a similar analysis indicates that
in the near horizon approximation, we have $f(u) = d(r_H/r_t)\alpha\sim d\alpha$.} Thus we can write in this case,
$f(u) = (4\pi T/r_t)\alpha$. First let us record the expression for $l$, given by
\begin{equation}
l = \sum_{n=0}^{\infty}\left(\frac{r_H}{r_t}\right)^n
\frac{\Gamma \left(\frac{n}{4}+\frac{3}{4}\right) \Gamma
   \left(n+\frac{1}{2}\right)}{2 r_t \sqrt{d} \Gamma
   \left(\frac{n}{4}+\frac{5}{4}\right) \Gamma (n+1)}
\label{lsadsnh}
\end{equation}
In this case, we find that the RT volume is
\begin{equation}
V = \sum_{n=0}^{\infty}\sum_{m=0}^{\infty}Lr_t\left(\frac{r_H}{r_t}\right)^{m+n}\left(T_1+ T_2 + T_3 + \cdots\right)
\end{equation}
where now we have 
\begin{eqnarray}
T_1 &=& \frac{2 \delta^{1+m+n}  \Gamma \left(m+\frac{1}{2}\right) \Gamma
   \left(n+\frac{1}{2}\right)}{3\pi (m+3) (m+n+1) \Gamma (m+1) \Gamma (n+1)} \nonumber\\
 T_2 &=& -\frac{\Gamma \left(\frac{m}{4}+\frac{3}{4}\right) \Gamma
   \left(m+\frac{1}{2}\right) \delta ^{n-2} \Gamma \left(n+\frac{1}{2}\right)}{6
   \sqrt{\pi }(n-2) \Gamma \left(\frac{m}{4}+\frac{5}{4}\right) \Gamma (m+1)
   \Gamma (n+1)}\nonumber\\
 T_3 &=&  \frac{\Gamma \left(m+\frac{1}{2}\right) \Gamma \left(n+\frac{1}{2}\right) \Gamma
   \left(\frac{m}{4}+\frac{n}{4}+\frac{1}{4}\right)}{6 \sqrt{\pi }(n-2) \Gamma
   (m+1) \Gamma (n+1) \Gamma \left(\frac{m}{4}+\frac{n}{4}+\frac{3}{4}\right)}
 \label{termssadsnh}
 \end{eqnarray}
Essentially the near horizon approximation has produced a similar expression as that in eq.(\ref{termssads}), with 
with slightly different numerics. In eq.(\ref{termssadsnh}), we see that the $n \leq 2$ terms give divergences. While the $n=1$ term
in $T_2$ of eq.(\ref{termssadsnh}) gives a divergence $\sim \delta^{-1}$, the $n=2$ piece in $T_3$ of that equation is undefined. 
Although $T_3$ is independent of $\delta$, we expect a $\log(\delta)$ divergence in that piece, and explicitly check this below. 
In the generic case, it should be then clear that the near horizon expansion is expected to produce spurious divergences 
$\sim \delta^{2-d}, \cdots \log(\delta)$ over and above the genuine $\delta^{1-d}$ divergence. 

To illustrate this explicitly, we will work out the $n=0$, $1$ and $2$ terms separately, and begin the series in $n$ from $n=3$. These terms are
found to be
\begin{eqnarray}
V_{n=0,d=3} &=& \sum_{m=0}^{\infty} Lr_t\left(\frac{r_H}{r_t}\right)^m\left(\frac{\Gamma \left(m+\frac{1}{2}\right) \Gamma \left(\frac{m+3}{4}\right)}{12
   \delta ^2 \Gamma (m+1) \Gamma \left(\frac{m+5}{4}\right)}-\frac{\Gamma
   \left(m+\frac{1}{2}\right) \Gamma \left(\frac{m+1}{4}\right)}{12 \Gamma (m+1)
   \Gamma \left(\frac{m+3}{4}\right)}\right)\nonumber\\
V_{n=1,d=3} &=& \sum_{m=0}^{\infty} Lr_t\left(\frac{r_H}{r_t}\right)^m
\left(\frac{\Gamma \left(\frac{m}{4}+\frac{3}{4}\right) \Gamma
   \left(m+\frac{1}{2}\right)}{12 \delta  \Gamma
   \left(\frac{m}{4}+\frac{5}{4}\right) \Gamma (m+1)}-\frac{2^{-\frac{3
   m}{2}-2}\pi \Gamma \left(m+\frac{1}{2}\right)}{3 \Gamma
   \left(\frac{m}{4}+1\right)^2 \Gamma \left(\frac{m}{2}+\frac{1}{2}\right)}\right)\nonumber\\
 V_{n=2,d=3} &=&  -\sum_{m=0}^{\infty} Lr_t\left(\frac{r_H}{r_t}\right)^m
\frac{3 \log (\delta ) \Gamma \left(\frac{m}{4}+\frac{3}{4}\right) \Gamma
   \left(m+\frac{1}{2}\right)}{48 \Gamma \left(\frac{m}{4}+\frac{5}{4}\right)
   \Gamma (m+1)}
\end{eqnarray}
Hence, we have in the limit $r_H \to r_t$, comparing with eq.(\ref{lsadsnh}), and labelling generic numerical factors as $a_i$,
\begin{equation}
V_{n=0} + V_{n=1} + V_{n=2} \sim Lr_t l\left(a_1 + a_2\frac{1}{\delta^2} + a_3\frac{1}{\delta} + a_4\log(\delta) + \frac{a_5}{lr_t})\right),~~d=3
\end{equation}
Now we note that $T_1$ produces a convergent result (as can be checked with the comparison test mentioned before) and we will discard this term. 
When summed from $n=3$, the $T_2$ and $T_3$ terms yield\footnote{We could not explicitly compute the sum over $T_3$ starting
from $n=3$. However, in the asymptotic limit, it has the same behaviour as the third term of eq.(\ref{termssads}) with $d=3$, and hence 
we conclude that this should go as $l$ apart from an additive convergent piece.}
\begin{equation}
V_{n\geq 3,d=3} \sim Lr_tl\left(a_6 + a_7\delta + \frac{a_8}{lr_t}\right)
\end{equation}
As mentioned before, comparing with the exact results of eq.(\ref{finalvsads}), we see that in $d=3$, 
spurious divergences of $1/\delta$ and $\log(\delta)$ have appeared in the near horizon computation, which are absent otherwise (the
$\delta^{-2}$ piece of course goes away when we subtract the pure AdS contribution). Similar results are expected to hold in
generic $d$ dimensions, leading to spurious divergences $\sim \delta^{-d+2}, \cdots \log(\delta)$. 
The same feature appears in the calculation of the entanglement entropy as well, as can be checked. With the inclusion of
a chemical potential, one is forced to resort to a near horizon analysis in the large temperature limit (or the large charge limit for extremal black
holes) due to the mathematical complications of an exact treatment (see, e.g \cite{GS}), and as we will see later, these spurious 
divergences will occur there as well, in addition to some possibly new features. 

\section{Planar Non-Extremal RN-AdS Black Holes}

In this section, we use the techniques used in the previous section to compute the volume enclosed by the minimal surface for a strip-like 
entangling surface in case of Reissner Nordstrom black holes, whose metric with the AdS length scale set to unity is given in $d + 1$ dimensions by,
\begin{eqnarray}
ds^2 &=& -r^2f(r)dt^2+\frac{1}{r^2 f(r)}dr^2+\frac{r^2}{L_{AdS}^2}(d{\vec x}^2)\\
f(r) &=& 1-\frac{M}{r^d} + \frac{Q^2}{r^{2d-2}}
\label{RN-AdSd}
\end{eqnarray}
Here, we will limit ourselves to the limiting regimes of high and low temperatures. Essentially, we will need to compute the two integrals of eq.(\ref{mainvl}) in these
two regimes. After solving for $M$ in terms of the horizon radius $r_H$, and noting that the Hawking temperature 
\begin{equation}
T = \frac{1}{4\pi}\left(dr_H - (d-2)Q^2r_H^{3-2d}\right)
\label{tempRN-AdS}
\end{equation} 
we define two quantities $\epsilon = r_H/r_t$ and $\alpha = \frac{r_t}{r_H} - u$, and write to equivalent
forms for $f(u)$ in terms of $\epsilon$ and $\alpha$ which can be used in the low and high temperature regimes $r_H \ll r_t$ and $r_H \sim r_t$
respectively. For example, in $d=3$, one has
\begin{equation}
f(u)_{d=3} = 1-\frac{u^3 \epsilon ^3
\left(r_H^4+Q^2\right)}{r_H^4}+\frac{Q^2 u^4 \epsilon ^4}{r_H^4}
\label{fulowt}
\end{equation}
\begin{equation}
f(u)_{d=3} = \frac{\left(3r_H^4 - Q^2\right)}{r_H^3r_t}\alpha - \frac{3\left(r_H^4 - Q^2\right)}{r_H^2r_t^2}\alpha^2
- \frac{\left(r_H^4 - 3Q^2\right)}{r_Hr_t^3}\alpha^3 + \frac{Q^2}{r_t^2}\alpha^4~
\label{fuhight}
\end{equation}
When $\epsilon$ is small, we can use eq.(\ref{fulowt}) and retain up to ${\mathcal O}(\epsilon^3)$ in that equation. In the other limit, 
when $r_H \sim r_t$, we can use eq.(\ref{fuhight}) in which we retain up to ${\mathcal O}(\alpha)$. For arbitrary $d$, 
the expression for $f(u)$ in the low temperature limit is 
\begin{equation}
f(u) = 1 - (u\epsilon)^d\left(1 + \frac{Q^2}{r_H^{2d-2}}\right) + Q^2\left(\frac{u\epsilon}{r_H}\right)^{2d-2}
\end{equation}
In the opposite limit, for general $d$, we have up to first order in $\alpha = \frac{r_t}{r_H} - u$,
\begin{equation}
f(u) = \frac{4\pi T}{r_t}\alpha + {\mathcal O}(\alpha^2)
\end{equation}
Let us first discuss this latter case. We see here that this is the same as the near horizon case for SAdS black holes that we studied
towards the end of the last section, see eq.(\ref{sadsnhmetric}) and the discussion after this equation. The analysis hence need not be 
repeated and the structure of the RT volume will be similar to that case. In particular, there might be spurious divergences 
over and above the $\sim \delta^{1-d}$ one, as an artefact of the near horizon expansion. 

In the low temperature limit, using eq.(\ref{fulowt}) in eq.(\ref{mainvl}), we obtain the finite part of the volume as 
\begin{equation}
V_{finite,~d=3} =-\frac{Lr_t \Gamma\left(\frac{1}{4}\right)^2 }{4 \sqrt{2 \pi }}
+\frac{L r_t\epsilon ^3
   \left(r_H^4+Q^2\right)}{4 r_H^4}+ {\mathcal O}(\epsilon^6)
   \label{lowtvol}
\end{equation}
We will compare this with the corresponding expression for the finite part of the RT area, which can be shown to be given in this limit by 
\begin{equation}
A_{finite,~d=3} = -\frac{2Lr_t \sqrt{\pi }\Gamma \left(\frac{3}{4}\right)}{\Gamma
   \left(\frac{1}{4}\right)} + \frac{L r_t \pi \epsilon ^3 \left(r_H^4+Q^2\right)}{4 r_H^4} + {\mathcal O}(\epsilon^6)
\label{lowtEE}
\end{equation}
Comparing eq.(\ref{lowtvol}) and eq.(\ref{lowtEE}), we see that these are the entirely similar expressions modulo the numerical factors. 
We record here the singularity structure of the two quantities. Whereas for the volume, we obtain
\begin{equation}
V_{singular,~d=3} = \frac{L}{\delta^2} \left(\frac{\pi  \epsilon ^3 r_t \left(r_H^4+Q^2\right)}{16
   r_H^4}+\frac{\sqrt{\pi } \Gamma \left(\frac{3}{4}\right) r_t}{\Gamma
   \left(\frac{1}{4}\right)}\right)
\end{equation}
the singular part of the RT area is 
\begin{equation}
A_{singular,~d=3} = \frac{2Lr_t}{\delta}
\end{equation}
Entirely similar results are expected in general dimensions, and we can directly use the result of \cite{SandipanP} to write the finite part of the
holographic complexity in any dimension by simply emulating eq.(3.17) of their paper. Of course, one can perform a more sophisticated 
analysis for non extremal RN-AdS black holes for the RT volume, in lines of what was done in \cite{SandipanP} for the entanglement 
entropy, by identifying various regimes in the space of the temperature and chemical potential. However, from our discussion, it should
be clear that this will be similar to the EE, modulo numerical coefficients and the details might be of limited interest.   

\section{Planar Extremal RN-AdS Black Holes}

According to the gauge-gravity duality, extremal black holes correspond to ground states of the boundary conformal field theory. From the metric
of eq.(\ref{RN-AdSd}), 
setting the Hawking temperature $T=0$, one obtains the charge parameter in terms of the radius of the horizon as,
\begin{equation}
Q = \sqrt{\frac{d}{d-2}}r_H^{d-1}
\end{equation}
which implies that for the extremal black hole, the lapse function can be written as,
\begin{equation}
f(r) = 1 -\frac{2(d-1)}{d-2}\left(\frac{r_H}{r}\right)^d + \frac{d}{d-2}\left(\frac{r_H}{r}\right)^{2d-2}
\end{equation}
We will consider two cases here, $r_h \ll r_t$ is the small charge case, and its opposite limit, $r_h \sim r_t$ is the large charge case. 
As before, in the former case, we use $r_H = \epsilon r_t$ to obtain 
\begin{equation}
f(u) = 1 - \frac{2(d-2)}{d-2}\left(u\epsilon\right)^d + \frac{d}{d-2}\left(u\epsilon\right)^{2d-2}
\label{extlowt}
\end{equation}
The above expression can be used in the small charge case, by retaining up to ${\mathcal O}(\epsilon^3)$ terms. In the large charge
regime, defining $\alpha = r_t/r_H - u$, we obtain up to third order in $\alpha$, 
\begin{equation}
f(u) = d(d-1)\left(\frac{r_H}{r_t}\right)^2\alpha^2 - \frac{d}{3}(5-8d + 3d^2)\left(\frac{r_H}{r_t}\right)^3\alpha^3 + \cdots
\label{exthight}
\end{equation}
In the near horizon approximation we may retain terms up to ${\mathcal O}(\alpha^2)$.

We will first understand the large charge regime. Here, by retaining the first term of eq.(\ref{exthight}), we obtain
$f(u)^{-1/2} \sim (r_t/r_H)\alpha^{-1}$, in all dimensions, apart from an unimportant dimension dependent multiplicative factor. 
We begin by first recording the expression for $l$, and it turns out to be 
\begin{equation}
l=\frac{d\sqrt{\pi}}{r_t d(d-1)^{\frac{3}{2}}}\sum_{n=0}^{\infty}
\left(\frac{r_H}{r_t}\right)^n\frac{\Gamma \left(\frac{d+n}{2 (d-1)}\right)}{\Gamma \left(\frac{2 d+n-1}{2
   (d-1)}\right)}
\label{lextr}
\end{equation}
This expression again has a formal divergence, and it is easy to see that the divergent part is a PolyLog function, 
$\text{Li}_{\frac{1}{2}}(r_H/r_t)$. We note here that in the limit $r_H \to r_t$, $l \sim 1/\sqrt{1-r_H/r_t} + \mathcal{O}\sqrt{(1-r_H/r_t)}$, 
as can be seen seen by making a series expansion of the PolyLog. 

Now, following the methods outlined in the previous sections, we obtain using standard methods a simple formula up to
${\mathcal O}(\delta)$ : 
\begin{equation}
V = L^{d-2}r_t^{d-2}\sum_{m=0}^{\infty}\sum_{n=0}^{\infty}\left(\frac{r_H}{r_t}\right)^{m+n}r_t\left(T_1 + T_2 + T_3 \right)
\label{vextr}
\end{equation}
where we have denoted 
\begin{eqnarray}
T_1 &=& \frac{2 \delta ^{m+n+1}}{(d-1) d (d+m) (m+n+1)}~,\nonumber\\
T_2 &=& -\frac{\sqrt{\pi } \Gamma \left(\frac{d+m}{2 d-2}\right) \delta ^{-d+n+1}}{d
   ((d-2) d+1) (-d+n+1) \Gamma \left(\frac{2 d+m-1}{2 d-2}\right)}~,\nonumber\\
T_3 &=& \frac{\sqrt{\pi } \Gamma \left(\frac{m+n+1}{2 d-2}\right)}{d ((d-2) d+1)
   (-d+n+1) \Gamma \left(\frac{d+m+n}{2 d-2}\right)}
\end{eqnarray}
Now in order not to clutter the notation, we will choose a specific example, $d=3$. The analysis for any other dimension can be done parallely. 
For $d=3$, there are potential divergence at $n \leq 2$, and so we will restrict the summation in eq.(\ref{vextr}) from $(m,n) = (0,3)$ and
consider the cases $n=0$, $1$ and $2$ separately. For $n=0$, we find 
\begin{equation}
V_{n=0,d=3} = \frac{Lr_H}{24}\sqrt{\pi}\sum_{m=0}^{\infty}\left(\frac{r_H}{r_t}\right)^m
\left(\frac{\Gamma \left(\frac{m}{4}+\frac{3}{4}\right)}{\delta^2  \Gamma
   \left(\frac{m}{4}+\frac{5}{4}\right)}-\frac{\Gamma
   \left(\frac{m}{4}+\frac{1}{2}\right)}{\Gamma \left(\frac{m}{4}+1\right)}\right)
\end{equation}
Both these are formally divergent series with the same divergence as that of $l$ in eq.(\ref{lextr}) and we thus have, denoting
numerical factors of ${\mathcal O}(1)$ by $a_i$,
\begin{equation}
V_{n=0,d=3} = Lr_Hl\left(a_1 + \frac{a_2}{lr_H} + \frac{a_3}{\delta^2}\right)
\end{equation}
In an entirely similar way, we can calculate
\begin{equation}
V_{n=1,d=3} = bLr_Hl\left(a_4 + \frac{a_5}{\delta}\right),~~V_{n=2,d=3} = Lr_Hl\left(a_6\log(\delta) + \frac{a_7}{lr_H}\right)
\end{equation}
Now we move to the rest of the terms in eq.(\ref{vextr}), where we sum from $(m,n) = (0,3)$. 
It can be checked that the term
involving $T_1$ goes to zero as $\delta \to 0$, by carefully taking the limit. The term involving $T_2$ also goes to zero in
the limit of $\delta \to 0$, for $d=3$.  
We can see this by first evaluating the sum over $m$,  which yields a complicated expression involving Hypergeometric functions
of $r_H/r_t$. We expand the result around $r_H/r_t=1$, and then perform the summation over $n$. This latter sum, in the limit $x \to 1$
goes to zero as $\delta \to 0$ (the same result can be obtained by first performing
the $n$ summation). Finally, we come to the term $T_3$. It might seem that this is a divergent term, since for large $m$ and $n$,
$T_3 \sim (m+n)^{-1/2}n^{-1}$ and while the $n$ sum converges for a finite $m$, the same is not true for the $m$ sum. 
In fact, this term might lead to interesting consequences. To see this, we first perform the
sum over $m$ and then expand the resulting series (involving products of Hypergeometric functions and gamma functions) around
$r_H/r_t = 1$. Then we obtain
\begin{equation}
\sum_{m=0}^{\infty} \left(\frac{r_H}{r_t}\right)^{m}T_3 = \frac{\pi  \left(4 n
   \left(\frac{r_H}{r_t}-1\right)-\frac{r_H}{r_t}-3\right)}{24 (n-2)
   \sqrt{1-\frac{r_H}{r_t}}} + {\mathcal S}
\label{summextr}
\end{equation}
where ${\mathcal S}$ is a combination of terms, all of which go as $n^{-3/2}$ for large $n$, and hence represent convergent sums. 
Keeping ${\mathcal S}$ aside, the sum over $n$ from $n=3$ can now be performed, and we obtain 
\begin{equation}
\sum_{m=0}^{\infty}\sum_{n=3}^{\infty} \left(\frac{r_H}{r_t}\right)^{m+n}T_3 = 
\frac{\pi  r_H^2}{ 24 r_t^2\sqrt{1-\frac{r_H}{r_t}}}\left[\left(11-\frac{7 r_H}{r_t}\right) \log
   \left(1-\frac{r_H}{r_t}\right)-\frac{4 r_H}{r_t}\right]
\end{equation}
This term then diverges as $l\log(1-r_H/r_t)$, i.e $\sim l\log(lr_t)$, in the near horizon limit. 
Hence, putting everything together, we finally have the expression for the RT volume in the large charge extremal RN-AdS black hole
in $d=3$ as
\begin{equation}
V_{d=3} \sim Llr_t\left(b_1 + \frac{b_2}{\delta^2} + \frac{b_3}{\delta} + b_4\log(\delta) + \frac{b_4}{lr_t} + b_5\log(lr_t)\right)
\end{equation}
The $\delta^{-d+1}$ divergence is removed by subtracting the pure AdS part. So in 
general $d$ dimensions, i.e there should be $\delta^{-d+2} \cdots \delta^{-1}$ along
with a $\log(\delta)$ divergence, which are spurious, as we have mentioned before. 
Note that the artefact of the near horizon approximation was only to produce extra spurious divergences in terms of
the ultra violet cutoff. The $\log(lr_t)$ term is not dependent on this, and seems generic to all dimensions. The reason it appears is also
simple to see. For large $m$ and fixed $n$, $T_3 \sim m^{-1/2}$ and this produces the $(1-r_H/r_t)^{-1/2}$ factor in eq.(\ref{summextr}). However
in that equation, the term that goes as $1/n$ for large $n$ produces the logarithm. This is therefore a mathematical consequence of 
the double sum, and is possibly not due to the near horizon approximation. 

It is important to point out that such a term did not occur in the computation of the entanglement entropy. The calculation is straightforward,
and gives in $d=3$,
\begin{equation}
A_{d=3} = Lr_t\left(\frac{a_1}{\delta} + a_2\log(\delta) + a_3 + \sqrt{\frac{\pi }{6}}\sum_{n=2}^{\infty} +\sum _{n=2}^{\infty }\left(\frac{r_H}{r_t}\right)^n
 \frac{\Gamma
   \left(\frac{n-1}{4}\right)}{2 \Gamma \left(\frac{n+1}{4}\right)}\right)
\end{equation}
By a slight rewriting of the gamma functions, the sum in the above equation can be shown to have a divergent piece proportional to $l$ (in $d=3$)
and a finite part. 

Next, let us take the small charge case. By using eq.(\ref{mainvl}), it is seen that in this case, 
\begin{eqnarray}
V_{finite,d=3} &=& Lr_t\left[-\frac{\Gamma(\frac{1}{4})^2}{4\sqrt{2\pi}} + \left(\frac{r_H}{r_t}\right)^3 + {\mathcal O}\left(\frac{r_H}{r_t}\right)^6\right] 
+ {\mathcal O}(\delta)\nonumber\\
V_{singular,d=3} &=& 
\frac{L r_t}{\delta^2} \left(\frac{\pi}{4}  \left(\frac{r_H}{r_t}\right)^3 -\frac{\sqrt{\pi } \Gamma
   \left(-\frac{1}{4}\right)}{16 \Gamma \left(\frac{5}{4}\right)}\right)
\end{eqnarray}
That the above equation is entirely similar to the entanglement entropy can be seen from eq.(4.7) of \cite{SandipanP} and we will not
discuss this further. The same situation occurs for generic $d$, and the conclusion is that the RT volume behaves in exactly the same
way as the entanglement entropy modulo numerical factors which are easy to work out. 

\section{Spherical RN-AdS Black Holes}

In this section we consider the Reissner-Nordstrom black holes in four dimensions with spherical topology, whose line element is,
\begin{eqnarray}
ds^2 &=& -f(r)dt^2 + \frac{dr^2}{f(r)}+r^2\left(d\theta^2 + \sin(\theta)^2 d\phi^2 \right)\\
f(r) &=& 1-\frac{M}{r}+\frac{Q^2}{r^2}+\frac{r^2}{L_{AdS}^2}
\end{eqnarray}
With the AdS length set to unity, the Hawking temperature of the black hole is  given by,
\begin{equation}
T = \frac{3r_h^4 + r_h^2 - Q^2}{4\pi r_h^3} = \frac{3 r_h^2+1-\Phi^2}{4\pi r_h}
\end{equation}
where $\Phi = \frac{Q}{r_h}$ is the potential.
The thermodynamics of this system has been extensively studied in both the fixed charge and fixed potential ensembles. It may be observed that the system undergoes a phase transition at the critical charge $Q_c = 1/6$. Therefore, we study the behaviour of the holographic complexity in the neighbourhood of this phase transition. 

We will choose the entangling region to be a spherical cap defined by $\theta \leq \theta_0$. The bulk parametrisation is 
defined by $r \equiv r(\theta)$, since we have rotational symmetry. The area functional is written as,
\begin{equation}
\mathcal{A} = 2\pi \int_{0}^{\theta_0}r \sin(\theta) \sqrt{\frac{r'(\theta)}{f(r)}+r^2}d\theta
\end{equation} 
The minimal surface is obtained by solving the equations of motion (which are cumbersome to write here), resulting from the area functional with the boundary conditions $r(\theta_0) \rightarrow \infty$ and $r'(0)=0$. 
 \begin{figure}[t!]
 \centering
 \subfigure[]{
 \includegraphics[width=2.5in,height=2.3in]{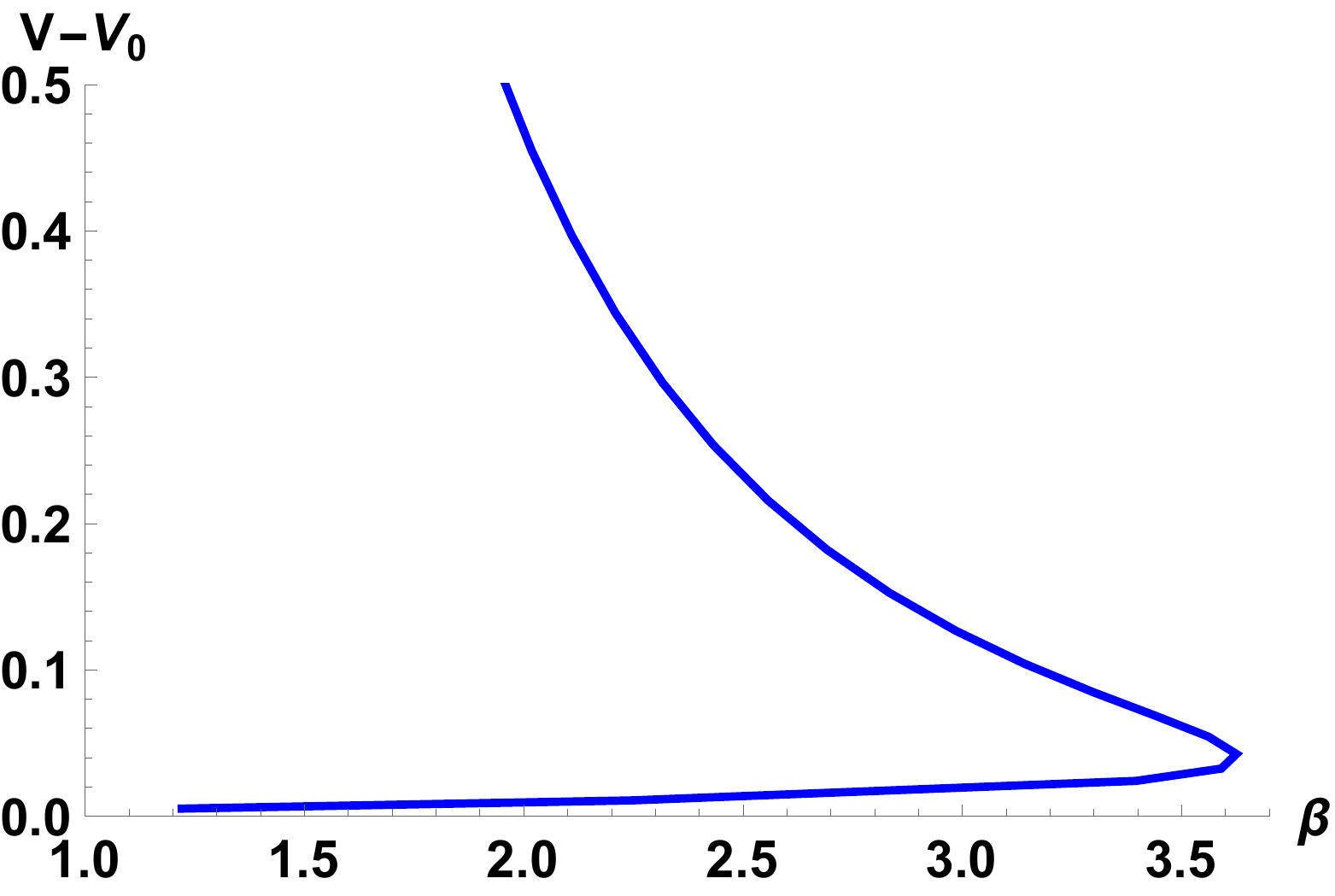}
 \label{fig1} } 
 \subfigure[]{
 \includegraphics[width=2.5in,height=2.3in]{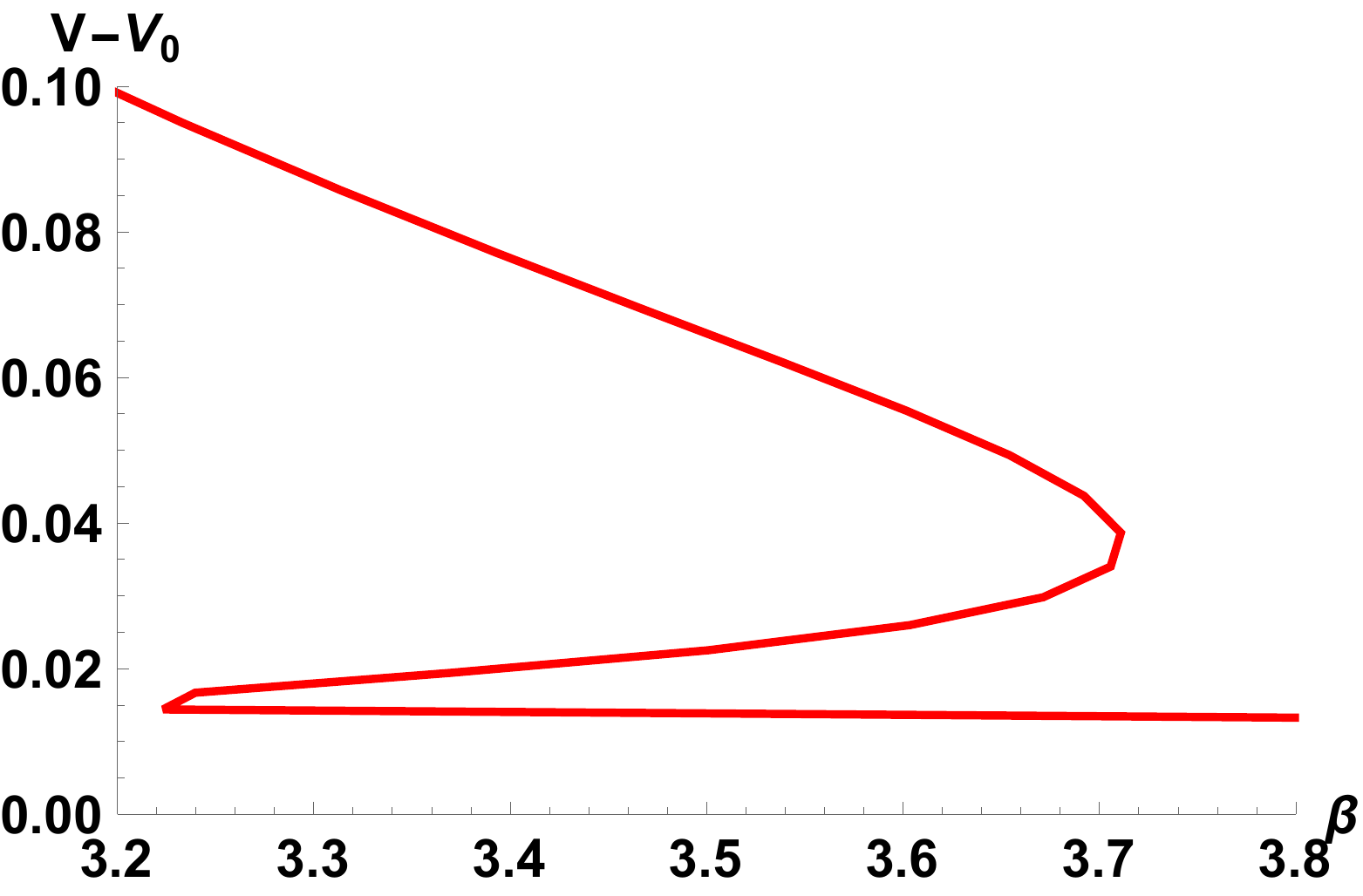}
 \label{fig2} }
  \caption{\small {In (a) the renormalized complexity is plotted for fixed charge ensemble with $Q = 0$ and the
opening angle $\theta_0 \approx 0.01$ In (b) we have fixed $Q = 1/6-0.005$ and the opening angle $\theta_0 \approx 0.01$.}}
 \end{figure}
 \begin{figure}[t!]
 \centering
 \subfigure[]{
 \includegraphics[width=2.5in,height=2.3in]{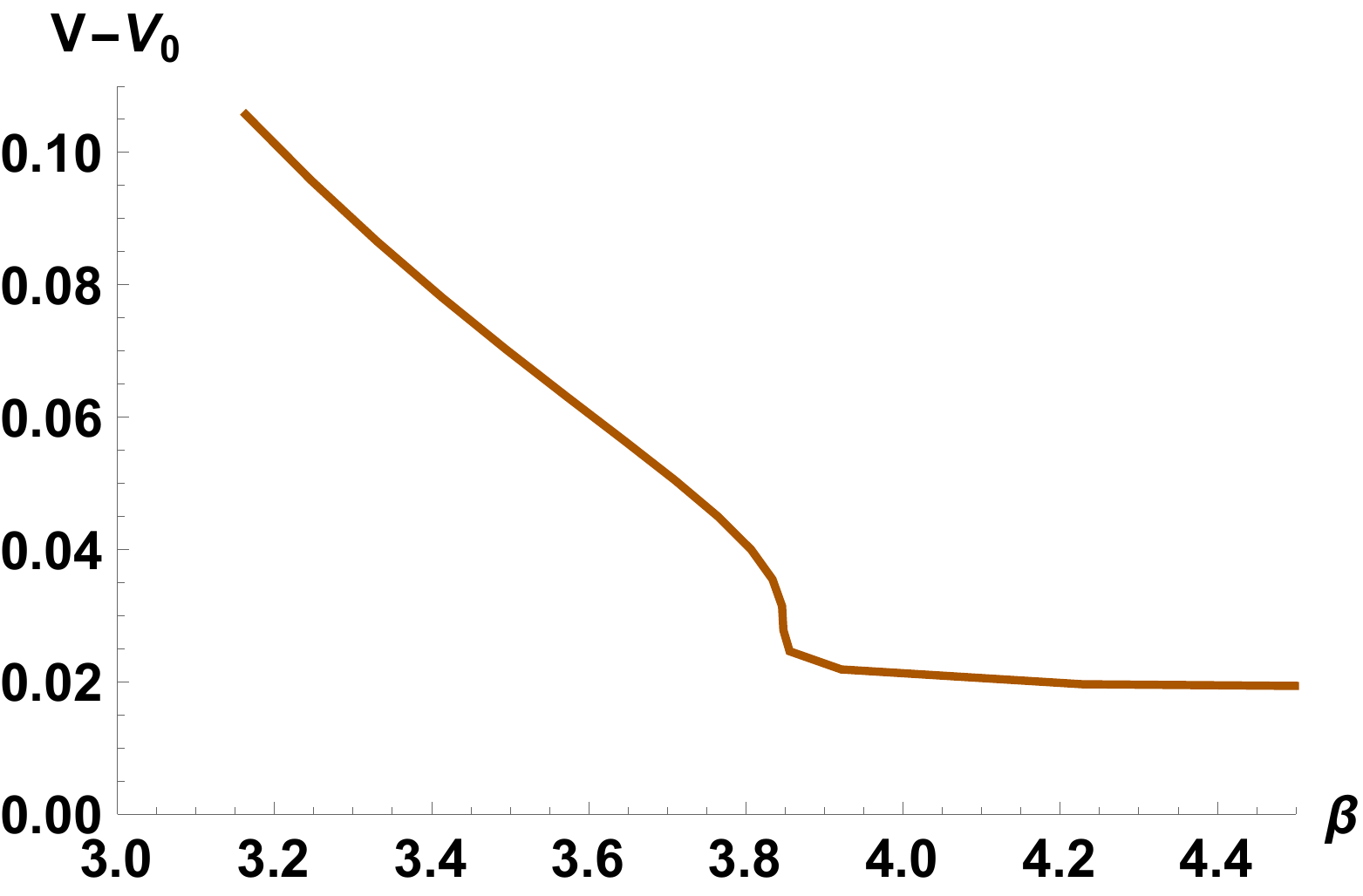}
 \label{fig3} } 
 \subfigure[]{
 \includegraphics[width=2.5in,height=2.3in]{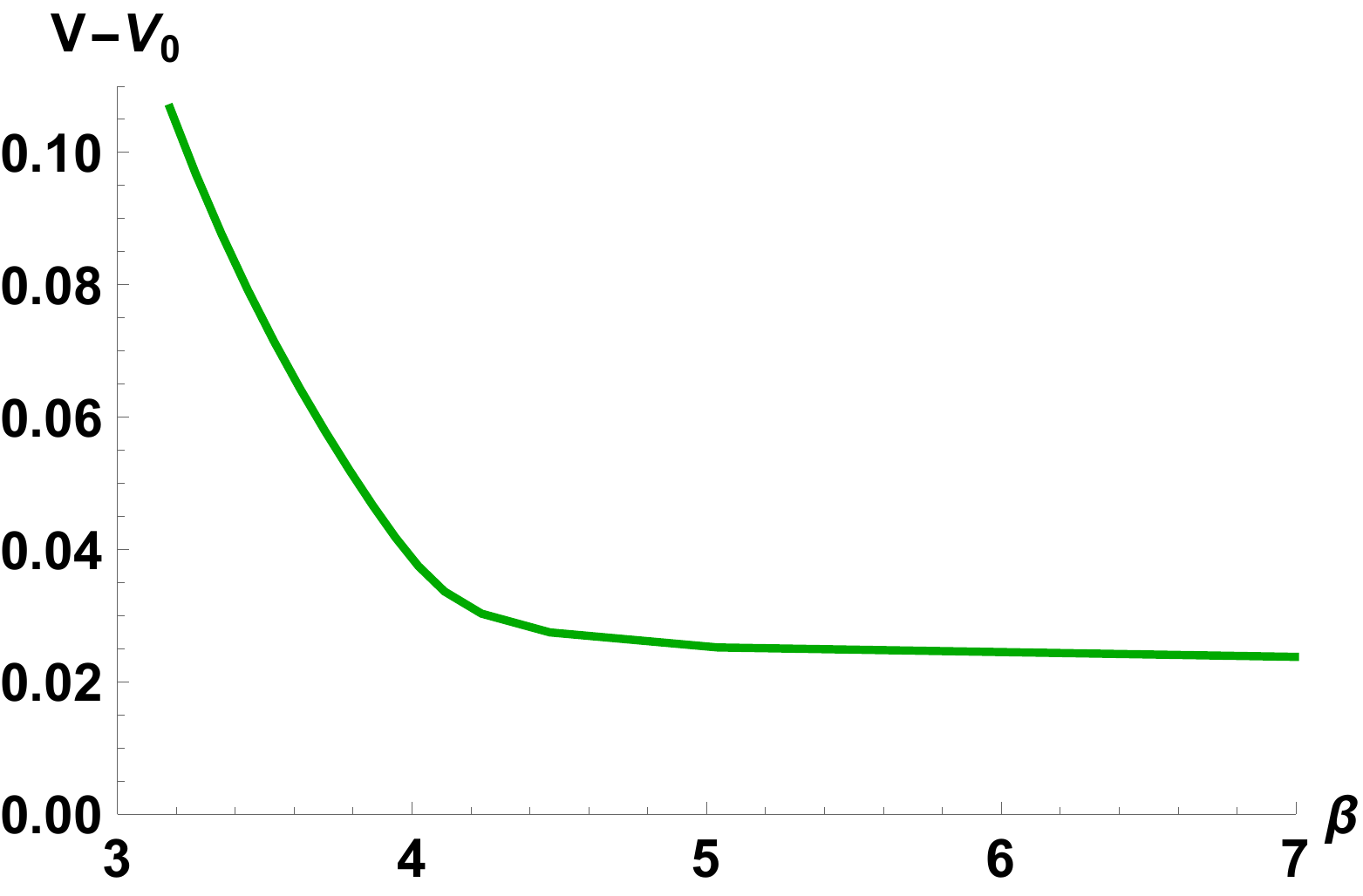}
 \label{fig4} }
  \caption{\small {In (a) the renormalized complexity is plotted for fixed charge ensemble with $Q = 1/6$ and 
the opening angle $\theta_0 \approx 0.01$ In (b) we have fixed $Q = 1/6+0.005$ and the opening angle $\theta_0 \approx 0.01$.}}
 \end{figure}
The volume enclosed by the bulk minimal surface is given by,
\begin{equation}
V = 4\pi \int_{0}^{\theta_0}d\theta~\sin(\theta)\int_{r_0 (\theta)}^{\infty} dr \frac{r^2}{\sqrt{f(r)}}
\end{equation}
where, the function $r_0 (\theta)$ is the solution to the equation of motion. Since the volume is divergent, we shall regularize it by subtracting the 
contribution due to pure global AdS. We show the behaviour of complexity as a function of the inverse temperature $\beta$. 
A word about the methods to obtain the RT volume from the above equation is now in order. We are confronted with two difficulties immediately. 
Firstly, the integral over $r$ can only be performed numerically, and secondly, we do not have an analytic form for the solution of the equation 
of motion $r_{0}(\theta)$. Therefore, we adopt the following strategy. Note that the integral over $\theta$ is performed from $0$ to $\theta_0$ 
(though in practice, we integrate from some cut-off $\delta$ up to $\theta_0$). We thus divide the range $0 \leq \theta \leq \theta_0$ into (say) 
$N$ points. The numerical solution $r_0 (\theta)$ is then used to determine the value of the $r$-integral for every one of these $N$ points. With 
the data in hand now of the $N$ points and the value of the $r$-integral for each of them, we can use interpolation to construct numerically the 
result of the $r$-integral as a function of $\theta$.\footnote{This procedure can be quite simply implemented in Mathematica by using the 
command \textit{Interpolation}.} The $r$-integral having been evaluated as a function of $\theta$, we can now numerically integrate the 
$\theta$-integral to obtain the desired result.

In figs.(\ref{fig1}) and fig(\ref{fig2}), we show the behaviour of the renormalised volume as a function of the inverse 
temperature $\beta = 1/T$ for the SAdS
and RN-AdS black holes in four dimensions, with a spherical entangling surface, respectively. In the latter case, we have chosen the charge
to be close to the critical charge, which for four dimensional RN-AdS black holes is $Q_c = 1/6$. In figs.(\ref{fig3}) and (\ref{fig4}), we show
the same quantities at criticality, and beyond criticality, respectively. It is seen that the RT volume behaves in the same way as the
entanglement entropy of \cite{Johnson} (see section 5 of that paper), and expectedly captures the behaviour of
the Davies transition, etc. In these figures, $V_0$ represent the contribution of a corresponding pure AdS.  

\section{Discussions and Conclusions}

In this paper, we have performed a systematic analysis of the high and low temperature behaviour of subregion holographic complexity
proposed in \cite{Alishahiha}. Initial work on the topic was reported in \cite{Omar}, and this work presents results that generalises 
analytically the ones reported in the latter. 

In this work, we have analysed the high and low temperature behaviour of the RT volume, which is conjectured to be the holographic
dual of subregion complexity (i.e complexity of a mixed state), in three situations, namely the $d+1$ dimensional SAdS, RN-AdS and
extremal AdS black holes. We have derived the necessary formulas for a strip geometry and related this volume to the strip length,
in all situations. Although for convenience, some of the results were presented for $d=3$, we have commented upon generic $d$ that
is easy to understand, following the methods that we have used. We have also worked out the details of the complexity for 
a  specific four dimensional geometry, for the case of a spherical entangling surface. 

Our results show that the behaviour of the RT volume mimics that of the entanglement entropy, apart from numerical factors in 
all the low temperature examples. These factors might be relevant for the boundary theory. In the high temperature cases, an analysis
of the double summations that arise in evaluating this volume showed that there might be some deviations from the
result for the EE for the extremal black hole with large charge, and this should be investigated further. In this paper we restricted
ourselves to time independent geometries, and it would be of obvious interest to extend our results beyond this. It might
also be interesting to analyse the behaviour of the RT volume for other classes of theories with hyperscaling violation. We leave
this to a future publication.

\end{document}